\documentstyle[12pt]{article}
\newcommand{\p}[1]{(\ref{#1})}
\newcommand{\cp}{\mbox{$\cal P$}}
\newcommand{\e}{\eta}
\newcommand{\be}{\begin{equation}}
\newcommand{\bea}{\begin{eqnarray}}
\newcommand{\ee}{\end{equation}}
\newcommand{\eea}{\end{eqnarray}}

\begin{document}
\setcounter{page}0
\renewcommand{\thefootnote}{\fnsymbol{footnote}}
\thispagestyle{empty}
{\hfill  Preprint JINR E2-98-303}\vspace{1.5cm} \\
\begin{center}
{\large\bf
On a different BRST constructions for a given Lie algebra
}\vspace{0.5cm} \\
A. Pashnev\footnote{E-mail: pashnev@thsun1.jinr.dubna.su}\\
and M. Tsulaia\footnote{E-mail: tsulaia@thsun1.jinr.dubna.su}
\vspace{0.5cm} \\
{\it JINR--Bogoliubov Theoretical Laboratory,         \\
141980 Dubna, Moscow Region, Russia} \vspace{1.5cm} \\
{\bf Abstract}
\end{center}
\vspace{1cm}

The method of the BRST quantization is considered for the system
of constraints, which form a Lie algebra. When some of the
Cartan generators do not imply any conditions
on the physical states, the system
contains the first and the second class constraints.
After the introduction auxiliary bosonic degrees of freedom
for these cases, the corresponding BRST charges with the nontrivial
structure of nonlinear terms in ghosts are constructed.

%\vspace{0.5cm}
\begin{center}
{\it Submitted to Proceedings of the Conference
``Supersymmetry and Quantum Symmetries"
dedicated to the memory of V.I. Ogievetsky}
\end{center}

\newpage\renewcommand{\thefootnote}{\arabic{footnote}}
\setcounter{footnote}0\setcounter{equation}0
\section{Introduction}

The  BRST quantization procedure  for a  system of the first
class constraints is straightforward. By the definition, the
first class constraints
form a closed algebra with respect to  the commutators
(the Poisson brackets).
For simplicity we  consider only linear algebras -- Lie algebras
of constraints.

More general systems  include the second class constraints as well,
whose commutators contain terms which are nonzero on mass shell
(on the subspace where all constraints vanish).
 In the simplest
cases  these terms are a numbers or central charges, but sometimes,
they are operators which act
nontrivially on the space of the physical states.
Moreover, the commutators between these
 operators and the constraints can be nontrivial. In some cases
the total system of the constraints and the
operators mentioned above form a Lie algebra.

So, in such cases we have a system of operators which form a Lie algebra,
but the physical meaning of different operators is different. Some of them
play the role of constraints and annihilate the physical states, others
are nonzero and simply transform the physical states into other ones.
It means, that in the BRST approach for the description of the
corresponding physical system we can not use the standard BRST charge
for the given Lie algebra.
Instead, we have to construct the nilpotent BRST
charge in  a manner, that
some of the operators play the role of the first class constraints,
others are second class constraints and the others
do not  imply any conditions on the physical
space of the system.

In this paper we demonstrate the possibility of a different
BRST constructions for the system of generators,
which form a given Lie
algebra and have different physical meaning.
In the second part we describe some algebraic approaches, leading
to the description of massless (or massive)
irreducible representations of the Poincare
group in any dimensions. As the simplest example we show,
that the same algebra of operators
leads to either massless, or massive spectra
in the cases when physical meaning of some
operators is different.
In the third part we discuss the general method of the BRST
quantization, when some of the Cartan generators are excluded
from the total system of constraints.
In the fourth part we describe the construction of auxiliary representations
of the algebra  by means of the
Gel'fand--Tsetlin method. In the Section 5 we give the simple
example.

\setcounter{equation}0\section{The description of constraints}
In order to  kill ghosts
the field theoretical lagrangians, describing irreducible Poincare
representations must possess some gauge invariance.
Along with the basic fields such lagrangians in general include
additional ones.
The role of these fields is to single out the irreducible
representation of the Poincare group.
Some of them are auxiliary, others can be gauged away.
After a gauge fixing and solving the equations of motion for auxiliary
fields one is left  with the only essential field, describing the
irreducible representation of the
Poincare group. This field
 corresponds to the Young tableaux with $k$ rows \vspace{1cm}\\

\begin{equation}
\begin{picture}(65,20)
\unitlength=1mm
\put(-20,20){\line(1,0){65.2}}
\put(-20,15){\line(1,0){65.2}}
\put(-20,10){\line(1,0){55}}
\put(-20,5){\line(1,0){45}}
\put(-20,0){\line(1,0){30}}
\put(-20,0){\line(0,1){20}}
\put(-15,0){\line(0,1){20}}
\put(-10,0){\line(0,1){20}}
\put(5,0){\line(0,1){20}}
\put(10.2,0){\line(0,1){20}}
\put(15,5){\line(0,1){15}}
\put(20,5){\line(0,1){15}}
\put(25,5){\line(0,1){15}}
\put(30,10){\line(0,1){10}}
\put(35.2,10){\line(0,1){10}}
\put(40,15){\line(0,1){5}}
\put(45.2,15){\line(0,1){5}}
\put(-19.5,17){$\mu_1$}
\put(-14.5,17){$\mu_2$}
\put(40,17){$\mu_{n_1}$}
\put(-19.5,12){$\nu_1$}
\put(-14.5,12){$\nu_2$}
\put(30,12){$\nu_{n_2}$}
\put(-19.5,2){$\rho_1$}
\put(-14.5,2){$\rho_2$}
\put(5,2){$\rho_{n_k}$}
\multiput(-7.5,17.5)(5,0){10}{\circle*{.5}}
\multiput(-7.5,12.5)(5,0){8}{\circle*{.5}}
\multiput(-17.5,7.5)(5,0){9}{\circle*{.5}}
\multiput(-7.5,2.5)(5,0){3}{\circle*{.5}}
\end{picture}
\label{young}
\end{equation}
and is described by
$
\Phi^{(k)}_{\mu_1\mu_2\cdots\mu_{n_1}, \nu_1\nu_2\cdots\nu_{n_2},
\cdots,\rho_1\rho_2\cdots\rho_{n_k}}(x)
$ which is the
  $n_1+n_2+\cdots+n_k$ rank tensor field
symmetrical with respect to the
permutations of each type of indices.
In addition, this field is subject to the following
system of equations, namely the mass shell and
transversality conditions
for each type of indices. In the massless
case we have
\begin{eqnarray}\label{mass}
&&p_{\mu}^2
\Phi^{(k)}_{\mu_1\mu_2\cdots\mu_{n_1}, \nu_1\nu_2\cdots\nu_{n_2},
\cdots,\rho_1\rho_2\cdots\rho_{n_k}}(x)
=0,\\
\label{trans1}
&&p_{\bf \mu}
\Phi^{(k)}_{{\bf \mu}\mu_2\cdots\mu_{n_1}, \nu_1\nu_2\cdots\nu_{n_2},
\cdots,\rho_1\rho_2\cdots\rho_{n_k}}(x)
=0,\\
&& \cdots\cdots\cdots\cdots\cdots\cdots\cdots\cdots\cdots\nonumber\\
\label{trans2}
&&p_{\bf \rho}
\Phi^{(k)}_{\mu_1\mu_2\cdots\mu_{n_1}, \nu_1\nu_2\cdots\nu_{n_2},
\cdots,{\bf \rho}\rho_2\cdots\rho_{n_k}}(x)
=0.
\end{eqnarray}
Further, all traces of the basic field must vanish:
\begin{eqnarray}
\label{trace1}
&&\Phi^{(k)}_{{\bf \mu\mu}\mu_3\cdots\mu_{n_1}, \nu_1\nu_2\cdots\nu_{n_2},
\cdots,\rho_1\rho_2\cdots\rho_{n_k}}(x)=0,\\
\label{trace2}
&&\Phi^{(k)}_{{\bf \mu}\mu_2\mu_3\cdots\mu_{n_1},{\bf \mu}
\nu_2\cdots\nu_{n_2},
\cdots,\rho_1\rho_2\cdots\rho_{n_k}}(x)=0,\\
&& \cdots\cdots\cdots\cdots\cdots\cdots\cdots\cdots\cdots\nonumber\\
\label{trace3}
&&\Phi^{(k)}_{\mu_1\mu_2\cdots\mu_{n_1}, \nu_1\nu_2\cdots\nu_{n_2},
\cdots,{\bf \rho\rho}\rho_3\cdots\rho_{n_k}}(x)=0.
\end{eqnarray}
The correspondence with a given Young tableaux implies, that
after symmetrization of all indices of one type with one index
of another type, the basic field vanishes,
for example
\begin{equation}
\label{sym}
\Phi^{(k)}_{\bf \{\mu_1\mu_2\cdots\mu_{n_1}, \nu_1\}\nu_2\cdots\nu_{n_2},
\cdots,\rho_1\rho_2\cdots\rho_{n_k}}(x)=0.
\end{equation}
To describe all irreducible representations of the Poincare group
simultaneously it is convenient to
introduce an auxiliary Fock space generated by
the creation and annihilation
operators $a^{i+}_\mu,a^j_\mu$
with Lorentz index $\mu =0,1,2,...,D-1$ and additional internal
index $i=1,2,...,k$. These operators satisfy the following
commutation relations
\be
\left[ a^i_\mu,a^{j+}_\nu \right] =-g_{\mu \nu}\delta^{ij},\;
g_{\mu \nu}=diag(1,-1,-1,...,-1),
\ee
where $\delta^{ij}$ is usual Cronecker symbol.

The general state of the Fock space
depends on the space-time coordinates $x_\mu$
\bea
|\Phi\rangle &=&\sum
\Phi^{(k)}_{\mu_1\mu_2\cdots\mu_{n_1}, \nu_1\nu_2\cdots\nu_{n_2},
\cdots,\rho_1\rho_2\cdots\rho_{n_k}}(x) \times \\
&&a^{1+}_{\mu_1}a^{1+}_{\mu_2}\cdots a^{1+}_{\mu_{n_1}}
a^{2+}_{\nu_1}a^{2+}_{\nu_2}\cdots a^{2+}_{\nu_{n_2}}
\cdots
a^{k+}_{\rho_1}a^{k+}_{\rho_2}\cdots a^{k+}_{\rho_{n_k}}
|0\rangle\nonumber
\eea
and the components
$\Phi^{(k)}_{\mu_1\mu_2\cdots\mu_{n_1}, \nu_1\nu_2\cdots\nu_{n_2},
\cdots,\rho_1\rho_2\cdots\rho_{n_k}}(x)$
are automatically symmetrical under the permutations of indices of
the same type \cite{OS} - \cite{LA2}.
The norm of states in this Fock space is not positively definite due to
the minus sign in the commutation relation (2.1) for the time components of
the creation and annihilation operators.
The transversality conditions \p{trans1}-\p{trans2}
for the components are equivalent to the following constraints on the
physical vectors of the Fock space
\begin{equation}
\label{trans3}
L^i|\Phi\rangle=0,
\end{equation}
where
\begin{equation}
\label{trans4}
L^i= a^i_\mu p_\mu.
\end{equation}
These operators along with their conjugates
\begin{equation}
\label{trans4}
L^{i+}= a^{i+}_\mu p_\mu.
\end{equation}
and mass shell operator $p^2_\mu$
form the following algebra with only nonvanishing commutator
\begin{equation}     \label{algebra1}
\left[L^i,L^{j+}\right]=-p^2_\mu\delta^{ij},
\end{equation}
This simple algebra was considered in \cite{OS} in the framework of
the BRST approach. The constraints are of the first class
and nilpotent BRST charge can be constructed without problems.
As a result
the description of mixed symmetry fields was obtained. However, all
these fields describe the reducible representations  of the Poincare group
  due to the absence of additional conditions
\p{trace1}-\p{trace3} and \p{sym} in the initial system of
the constraints.

On the other hand, the same algebra
of operators arises
in the case of massive particles. The only difference is that the
right hand side of the relation \p{algebra1} is now nonvanishing
operator. Instead, this operator can have different eigenvalues
$p^2_\mu=m_n^2$ for
the different physical states. This situation was
analyzed in \cite{PT2},
where corresponding BRST charge,
obviously different from the one for the
massless case, was constructed using the method of dimensional reduction.
 As an artifact of this method the construction automatically
includes some additional auxiliary variables.
So,  \p{algebra1} produces the nontrivial example
for the different  BRST constructions, corresponding
to different physical meaning of the generators ($p^2_\mu$ in our case).
In what follows we  describe the method of alternative
constructions of BRST charges which is valid not only for a simple
algebras like \p{algebra1}.

The tracelessness conditions
\p{trace1}-\p{trace3} correspond in the  Fock space to the
constraints
\begin{eqnarray}
\label{trace4}
L^{ij}|\Phi\rangle=0,
\end{eqnarray}
with
\begin{equation}
\label{trace5}
L^{ij}=a^i_\mu a^j_\mu,\;\;\;\;\;L^{ij+}=a^{j+}_\mu a^{i+}_\mu.
\end{equation}
while the symmetry properties \p{sym} follow from the
constraints
\begin{equation}
\label{sym2}
T^{ij}|\Phi\rangle=0, \quad \;\;i<j,
\end{equation}
having the  explicit form
\begin{equation}
\label{sym3}
T^{ij}=a^{i+}_\mu a^j_\mu,\;\;\;T^{ij+}=a^{j+}_\mu a^{i}_\mu=T_{ji}.
\end{equation}

The operators $L^{ij}, \;L^{ij+}$ ($i,j$ are arbitrary) and
$T^{ij},\; (i\neq j)$, along with the additional operators
\begin{equation}
\label{Cartan}
H^i= -T^{ii}+\frac{D}{2}=-a^{i+}_\mu a^i_\mu+\frac{D}{2},
\end{equation}
 form the Lie algebra $SO(k+1,k)$.
The rank of this algebra is $k$ and corresponding Cartan subalgebra
contains all operators $H^i$. One can choose
the operators $L^{11}$ and $T^{i,i+1}$ as $k$ simple roots.
 The positive and negative
roots are, correspondingly, $L^{ij},\;T^{rs},\;(1\leq r<s\leq k)$
and $L^{ij+},\;T^{rs},\;(1\leq s<r\leq k)$. It means, that the
conditions \p{trace1}-\p{trace3} and \p{sym} are equivalent
to annihilation of physical states in the total Fock space
by the positive roots of the Lie algebra $SO(k+1,k)$.

As it can be easily seen, the Cartan generators
\p{Cartan}  are strictly positive in the Fock space and therefore
the standard BRST charge has to be modified for the given
realization of the  $SO(k+1,k)$ algebra.

The BRST approach to the construction of the lagrangians, from which
all the equations \p{mass}-\p{sym} follow, is very powerful.
It automatically leads to appearance of all auxiliary fields
in the lagrangian.
In the massless case the BRST charge for the system of only
first class constraints, corresponding to the equations
\p{mass} - \p{trans2} was constructed in \cite{OS}.

The methods of such construction
were discussed in \cite{FS} - \cite{EM}.
With the help of additional variables
one can modify the second class constraints in such a way that they
become commuting, i.e. the first class. At the same time the number
of physical degrees of
freedom for both systems does not change if the number of
additional variables coincides with the number of second class
constraints.

On the other hand, the BRST charge for the second class constraints
in some cases can be constructed using the
method of dimensional reduction. In \cite{PT2} the system of massive
higher spins satisfying equations
\begin{equation}
(p_{\mu}^2-m^2)
\Phi_{\mu_1\mu_2\cdots\mu_{n_1}}(x)=0
\end{equation}
and
\begin{equation}\label{trans}
p_{\bf \mu}\Phi_{{\bf \mu}\mu_2\cdots\mu_{n_1}}(x)=0
\end{equation}
was described in the framework of the BRST approach. From the point of view
of $D$ dimensions, where constraint \p{trans} is of second class,
the $D+1$-st components of the creation
and annihilation operators appear in the consideration as additional
operators.
The corresponding BRST charge is nilpotent and has a very special
structure. In particular, the modified constraints have the algebra,
which is not closed. Nevertheless, the nontrivial structure
of trilinear terms in ghosts in the BRST charge compensates this
defect and makes the BRST charge to be nilpotent.
Another example of BRST charge for the system, including second class
constraints was obtained in \cite{PT3}. It reproduces some properties
of the BRST charge of \cite{PT2}: the algebra of modified constraints
is not closed and trilinear terms in ghosts are nontrivial as well.
Moreover, the BRST charge   contains terms up to the
seventh degree in ghosts.
In the next Section  we will
describe the simple method, which allows to
one to construct the nilpotent
BRST charges for a given Lie algebra,
when some generators are treated as the second class constraints.
As a particular case, this method reproduces the results
obtained in  \cite{PT2} and \cite{PT3}.

\setcounter{equation}0\section{The general method}
 In this section, we  describe the
method of the BRST construction, which leads to the desirable division
of the generators of a given Lie algebra
 into the first and second class constraints.
Let $H^i,\;(i=1,...,k)$ and $E^\alpha$
be the Cartan generators and root vectors
of the algebra with the following commutation relations
\begin{eqnarray}
\label{commutator}
&&\left[H^i,E^\alpha\right]=\alpha(i) E^\alpha,\\
&&\left[E^\alpha,E^{-\alpha}\right]=\alpha^i H^i,\\
&&\left[E^\alpha,E^{\beta}\right]=N^{\alpha\beta}E^{\alpha+\beta}.
\end{eqnarray}
Roots $\alpha(i)$ and parameters $\alpha^i,\; N^{\alpha\beta}$
are structure constants of the algebra in the Cartan - Weyl basis.
Our goal is to construct nilpotent BRST charge, which after quantization
 leads to the following conditions: all positive root vectors
$E^\alpha\;(\alpha>0)$ of
the algebra annihilate the physical states. Contrary, the operators
$H^i$ which form the Cartan subalgebra
may or may not be constraints, depending
on the physical nature of these operators.

The simplest case, when
all Cartan generators annihilate the physical states, is well known.
We introduce the set of anticommuting variables
$\e_i,\e_\alpha,$ $\e_{-\alpha}=\e_\alpha^+$, having ghost number one
and corresponding momenta
$\cp_i,\cp_{-\alpha}=\cp_\alpha^+,\cp_\alpha$, with the
commutation relations:
\begin{equation}
\{\e_i,\cp_k\}=\delta_{ik},\;\{\e_\alpha,\cp_{-\beta}\}=
\{\e_{-\alpha},\cp_\beta\}=\delta_{\alpha\beta}
\end{equation}
we define the ``ghost vacuum" as
\begin{equation}
\e_\alpha|0\rangle=\cp_\alpha|0\rangle =\cp_i|0\rangle=0
\end{equation}
for positive roots $\alpha$.
The BRST charge for the Cartan - Weyl decomposition of the algebra
has a standard form
\begin{eqnarray}
Q&=&\sum_i\e_i H^i+\sum_{\alpha>0}\left(\e_\alpha E^{-\alpha}+
\e_{-\alpha}E^{\alpha}\right)-
\frac{1}{2}\sum_{\alpha\beta}N^{\alpha\beta}
\e_{-\alpha}\e_{-\beta}\cp_{\alpha+\beta}
+        \nonumber      \\                    \label{brst1}
&&\sum_{\alpha>0,i}\{\alpha(i)\left(\e_i\e_\alpha\cp_{-\alpha}-
\e_i\e_{-\alpha}\cp_\alpha\right)+\alpha^i\e_\alpha\e_{-\alpha}\cp_i
\}
\end{eqnarray}
The physical states are then the cohomology classes of the BRST operator.

The quantization in this case is equivalent to the quantization
${\grave a}$ la Gupta - Bleuler, because physical states satisfy
equations $H^i|Phys\rangle=0$ and\\ $E^\alpha|Phys\rangle=0$
only for positive values of $\alpha$.

The situation changes when some of the Cartan operators
$H^i$, say $H^{i_l},\;l=1,2,...N$  are
nonvanishing from the  physical reasons. In this case the following
method can be used.

First of all we construct some auxiliary representation for the
generators $H^i,\; E^\alpha$ of the
algebra in terms of additional creation and annihilation operators.
The only condition for this representation is that it depends
on some parameters $h^{n}$. The total number of
these parameters is equal to  the
number of the Cartan generators, which are nonzero in the physical
sector. In what follows, we  consider the realizations of the
algebra with a linear dependence of the Cartan generators on these
parameters: ${\hat H}^{m}(h)=\tilde{H}^{m}+c^m_n h^{n}$, where
$c^m_n$ are  some constants.
The  $h^n$ dependence of other generators can be arbitrary.
In the next section
we describe the general method of construction of such representations.
Here we simply assume that they exist.

The next step is to consider the realization of the algebra
as a sum of "old" and "new" generators
$${\cal H}^i=H^i+{\hat{H}}^i(h) ,\quad
{\cal E}^\alpha=E^\alpha+{\hat E}^\alpha(h).$$
The BRST charge for the total system has the same form as
\p{brst1}, with modified generators:
\begin{eqnarray}
{\cal Q}&=&\sum_i\e_i {\cal H}^i+
\sum_{\alpha>0}\left(\e_\alpha {\cal E}^{-\alpha}+
\e_{-\alpha}{\cal E}^{\alpha}\right)-
\frac{1}{2}\sum_{\alpha\beta}N^{\alpha\beta}
\e_{-\alpha}\e_{-\beta}\cp_{\alpha+\beta}
+        \nonumber      \\                    \label{brst2}
&&\sum_{\alpha>0,i}\{\alpha(i)\left(\e_i\e_\alpha\cp_{-\alpha}-
\e_i\e_{-\alpha}\cp_\alpha\right)+\alpha^i\e_\alpha\e_{-\alpha}\cp_i
\}
\end{eqnarray}
The ghost variables $\eta_{i_l}$, correspond  to the set of
nonvanishing generators $H^{i_l}$ and therefore one needs to
remove the $\eta_{i_l}$ dependence
\begin{equation}
Q_{i_l}=\e_{i_l}\{
H^{i_l} +\tilde{H}^{i_l}+c^{i_l}_nh^n
+\sum_{\beta>0}\alpha(i_l)\left(\e_\beta\cp_{-\beta}-
\e_{-\beta}\cp_\beta\right)
\}.
\end{equation}
from the BRST  charge. For this purpose
consider an auxiliary $N$ - dimensional
space with coordinates $x_{i_l}$ and conjugated momenta $p^{i_l}$,
where $c^{i_l}_nh^n=p^{i_l}$:
\begin{equation}
\left[x_{i_l}, p^{i_n}\right]=i \delta_{i_l}^{i_n}.
\end{equation}
After the similarity transformation, which corresponds
to the dimensional reduction \cite{PT2}
\begin{equation}       \label{transformation}
\tilde{\cal Q}=e^{i\pi^{i_l} x_{i_l}} Q e^{-i\pi^{i_l} x_{i_l} },
\end{equation}
where
\begin{equation}
\pi^{i_l}=
H^{i_l} +\tilde{H}^{i_l}+
+\sum_{\beta>0}\alpha(i_l)\left(\e_\beta\cp_{-\beta}-
\e_{-\beta}\cp_\beta\right)
\end{equation}
the transformed BRST charge $\tilde{\cal Q}$ does not depend on the ghost
variables $\eta_{i_l}$. All parameters $p^{i_l}$ in the BRST charge
are replaced by the corresponding operators $\pi^{i_l}$.
The transformation \p{transformation}
does not change the nilpotency property of the BRST charge.
It means that the $\cp_{i_l}$ independent part $\tilde{\cal Q}_0$ of the
total charge $\tilde{\cal Q}$ is nilpotent as well.
Moreover, as a consequence of the nilpotency of $\tilde{\cal Q}$
all coefficients  at the corresponding
antighost operators  $\cp_{i_l}$ commute with $\tilde{\cal Q}_0$.
One can show that the quantization with the help of the BRST
operator $\tilde{\cal Q}_0$ will lead to the desirable reduced system of
constraints on the physical states.

\setcounter{equation}0\section{Construction of auxiliary representations
of the algebra}
Consider the highest weight representation of the algebra under
consideration with the highest weight vector $|\Phi\rangle$ annihilated
by the positive roots
\begin{equation}
E^\alpha|\Phi\rangle=0
\end{equation}
and being the proper vector of the Cartan generators
\begin{equation}
H^i|\Phi\rangle=h^i|\Phi\rangle.
\end{equation}
As it was  shown by Gelfand and Tsetlin \cite{GT}, each of the vectors
of the irreducible representation with a given highest weight can be
associated with the so called Gelfand-Tsetlin scheme. Corresponding
scheme for $U(k)$ algebra has the following form:

\begin{tabular}{|ccccccccc|}
$m_{1,k}$&&$m_{2,k}$&&&&$m_{k-1,k}$&&$m_{k,k}$\\
&$m_{1,k-1}$&&&.&&&$m_{k-1,k-1}$&\\
.&.&.&.&.&.&.&.&.\\
&&&$m_{12}$&&$m_{22}$&&&\\
&&&&$m_{11}$&&&&
\end{tabular}\vspace{0.5cm}\\

The first row of the scheme is defined by the highest weight components:
\begin{equation}
m_{i,k}=h^i.
\end{equation}
This row is fixed for a given irreducible representation of $U(k)$.
All other rows contain arbitrary numbers under the following conditions:
\begin{equation}\label{condition}
m_{ij}\geq m_{i,j-1}\geq m_{i+1,j},\;j=2,...,k;\;i=1,...,k-1.
\end{equation}
Any choice of this numbers, consistent with \p{condition},
corresponds to a fixed vector in the irreducible representation.
All these vectors are orthonormal.
The total number of $m_{i,j}$ coincides with the number of positive
roots of the algebra $U(k)$. This is true for any semisimple algebra.
It means, that all such vectors can be represented in an auxiliary
Fock space, generated by the oscillators $b_{i,j}, b^+_{i,j}$,
which are in one to one correspondence with numbers $m_{i,j}$.
So the vector, which corresponds to the Gelfand - Tsetlin scheme
given above has the following form
\begin{equation}
|\{m_{i,j}\}\rangle =\prod_{i,j}\frac{1}{(m_{i,j}-\kappa_{i,j})^{1/2}}
(b_{i,j}^+)^{(m_{i,j}-\kappa_{i,j})}|0\rangle,
\end{equation}
where $\kappa_{i,j}$ are some fixed parameters, connected with
the weights $h^i$. Having at hands all matrix elements of generators
\cite{GT} one can easily reconstruct the expressions for the generators
in terms of oscillators $b_{i,j}, b^+_{i,j}$. They will depend
on $k$ parameters $h^i$ and give the needed auxiliary representation
of the algebra. Below  we construct such representation
for the simplest example of
$SO(2,1)$ to illustrate the method of the BRST construction
described in the previous section.

\setcounter{equation}0\section{The simple example}
In this section we consider $SO(2,1)$ algebra of constraints.
The Fock space introduced in the Section 2 is spanned by only one oscillator.
This system describes higher spin irreducible
representations of the Poincare algebra corresponding to the Young
tableaux with only one row. The total system of generators includes
\begin{eqnarray}
\label{so21tr}
L_1&=& a_\mu p_\mu,\quad
L_{-1}= a^{+}_\mu p_\mu,\quad L_0=p^2_\mu,\quad
\left[L_1,L_{-1}\right]=-L_0,\\            \label{so21}
L_2&=&a_\mu a_\mu,\quad L_{-2}=a^+_\mu a^+_\mu,\quad
\left[L_2,L_{-2}\right]=-a^+_\mu a_\mu-\frac{D}{2}\equiv G_0,
\end{eqnarray}
where $D$ is the dimensionality of the space -- time.
The last three operators $L_2,\;L_{-2}$ and $G_0$ form an $SO(2,1)$
algebra.
The first three operators in \p{so21tr} transform as a
representation of this algebra and they can be
included in the BRST charge rather trivially. So, the main problem
is to construct the BRST charge for $SO(2,1)$ algebra under the
condition, that the operator $G_0$ is not a constraint in the
physical subspace, since
$G_0$ is positively definite in the whole Fock space.

Using the results of the previous Section one can easily  construct
the auxiliary representation of the algebra $SO(2,1)$ in terms
of one additional timelike oscillator $
\left[b,b^+\right]=-1
$:
\begin{eqnarray}
\hat{L}_2&=&\sqrt{h+b^+b}\;b,\nonumber\\
\label{aux}
\hat{L}_{-2}&=&b^+\;\sqrt{h+b^+b},\\
\hat{G}_0&=&\tilde{G}_0-h,\quad \tilde{G}_0=-2b^+b.
\end{eqnarray}

The BRST charge for modified generators
\begin{equation}
{\cal L}_{2}=L_{2}+\hat{L}_{2},\quad
{\cal L}_{-2}=L_{-2}+\hat{L}_{-2},\quad
{\cal G}_{0}=G_{0}+\hat{G}_{0}
\end{equation}
takes the following form
\begin{equation}
{\cal Q}=\e_0 {\cal G}_0+
\e_2 {\cal L}_{-2}+
\e_{-2}{\cal L}_{2}-
                    \label{brst2}
2\e_0\e_2\cp_{-2}+
2\e_0\e_{-2}\cp_2+\e_2\e_{-2}\cp_0
\end{equation}
As the result of similarity transformation  described in the Section 3,
the coefficient at ghost variable $\eta_0$
vanishes and the following replacement of
parameter $h$ takes place:
 \begin{equation}
h\rightarrow G_0-2b^+b + 2\cp_{-2} \e_2+2\e_{-2}\cp_2
\end{equation}
The resulting nilpotent BRST charge after removing of the antighost
variable $\cp_0$
looks as follows
\begin{eqnarray}
{\cal Q}_0&=&\e_2 \{{L}_{-2}+b^+\;
\sqrt{G_0-b^+b+2\cp_{-2}\e_2+2\e_{-2}\cp_2}\}+\\
&&\e_{-2}\{{L}_{2}+\sqrt{G_0-b^+b+2\cp_{-2}\e_2+2\e_{-2}\cp_2}\;b\}.
\nonumber
\end{eqnarray}
the inclusion of the constraints $L_0$, $L_1$ and $L_{-1}$
into the BRST charge is trivial. After the corresponding gauge fixing
and solving the equations of motion  for some of the auxiliary  fields,
one obtains the BRST quantized lagrangian, given in
\cite{F}.

\setcounter{equation}0\section{Conclusions}
In this paper we have demonstrated  the  various constructions of the
nilpotent BRST charges  for a given algebra
of constraints. The identification of the generators of the algebra
with the constraints on the physical states is model--dependent
and therefore after the quantization
 these BRST charges lead to the different
spectrum of the physical states.\vspace{.5cm}\\

\noindent {\bf Acknowledgments.}
This investigation has been supported in part by the
Russian Foundation of Fundamental Research,
grants 96-02-17634 and 96-02-18126,
joint grant RFFR-DFG 96-02-00180G,
and INTAS, grants 93-127-ext, 96-0308,
96-0538, 94-2317 and grant of the
Dutch NWO organization.


\begin{thebibliography}{99}
\bibitem{OS} S.Ouvry, J.Stern. Phys.Lett., B177 (1986) 335
\bibitem{LA1} J.M.F.Labastida. Phys.Rev.Lett., 58 (1987) 531
\bibitem{LA2} J.M.F.Labastida. Nucl.Phys., B322 (1989) 185
\bibitem{PT2} A. Pashnev, M. Tsulaia. Mod.Phys.Lett., A12 (1997) 861
\bibitem{FS} L.D. Faddeev, S.L. Shatashvili. Phys.Lett., B167 (1986) 225
\bibitem{BF} I.A. Batalin, E.S. Fradkin. Nucl.Phys., B279 (1987) 514
\bibitem{EM}  E.T. Egoryan, R.P. Manvelyan. Theor. Math.Phys.,
94 (1993) 241
\bibitem{PT3} A. Pashnev, M. Tsulaia. Mod.Phys.Lett., A13 (1998) 1853
\bibitem{GT} I.M.Gel'fand, M.L.Tsetlin. Dokl. Akad. Nauk SSSR.,
71 (1950) 825; ibid 71 (1950) 1017
\bibitem{F} C.Fronsdal. Phys.Rev., D18
 (1978) 3624
\end{thebibliography}
\end{document}